\documentclass[prb,english,twocolumn,amsmath,amssymb,superscriptaddress]{revtex4-2}
\usepackage{amsmath}
\usepackage{graphicx}
\usepackage{amssymb}
\usepackage{dsfont}
\usepackage{color}
\usepackage[bookmarks=true,colorlinks,linkcolor=blue,urlcolor=blue,citecolor=blue]{hyperref}

\usepackage{hyperref}
\usepackage{tikz}
\usetikzlibrary{calc}

\newcommand{\be}{\begin{equation}}
\newcommand{\ee}{\end{equation}}
\newcommand{\bea}{\begin{eqnarray}}
\newcommand{\eea}{\end{eqnarray}}
\newcommand{\ket}{\rangle}
\newcommand{\bra}{\langle}
\newcommand{\mc}{\mathcal}
\newcommand{\mb}{\mathbf}

\definecolor{darkorange}{rgb}{1, 0.55, 0.0}

\begin{document}
\title{Continuous phase transition from  a chiral spin state to collinear magnetic order in a zigzag chain with Kitaev interactions}

\author{Rafael A. Mac\^{e}do}
\affiliation{Departamento de F\'{i}sica Te\'{o}rica e Experimental, Universidade Federal do Rio Grande do Norte, Natal, RN, 59078-970, Brazil}
\author{Fl\'avia B. Ramos}
\affiliation{International Institute of Physics, Universidade Federal do Rio Grande do Norte, Natal, RN, 59078-970, Brazil}
\affiliation{Physics Department and Research Center OPTIMAS,
Technische Universit\"at Kaiserslautern, 67663 Kaiserslautern, Germany}
\author{Rodrigo G. Pereira}
\affiliation{Departamento de F\'{i}sica Te\'{o}rica e Experimental, Universidade Federal do Rio Grande do Norte, Natal, RN, 59078-970, Brazil}
\affiliation{International Institute of Physics, Universidade Federal do Rio Grande do Norte, Natal, RN, 59078-970, Brazil}

\begin{abstract}
 
Quantum spin  systems can break time reversal  symmetry by developing spontaneous magnetization or spin chirality. However, collinear magnets and chiral spin states are invariant under different symmetries, implying  that the  order parameter of one phase vanishes in  the other. We show how to construct one-dimensional anisotropic spin models that exhibit a ``Landau-forbidden'' continuous phase transition between such  states.  As a concrete example, we focus on a zigzag chain   with bond-dependent exchange  and six-spin interactions. Using a combination of exact solutions,  effective field theories, and numerical simulations, we show that the transition between the chiral and     magnetic phases  has an emergent U(1) symmetry. The excitations   governing the transition from the chiral phase can be pictured as mobile defects in a $\mathbb Z_2$ flux configuration which bind fermionic modes. We briefly discuss  extensions to two dimensions    and analogies with deconfined quantum criticality. Our results suggest new prospects for unconventional phase transitions involving  chiral spin states. 

\end{abstract}

\date\today

\maketitle

\section{Introduction}

The most familiar type  of  spontaneous symmetry breaking  in   quantum spin systems   is magnetic long-range order.  A paramount example is the N\'eel order in the  ground  state of the antiferromagnetic Heisenberg model on bipartite lattices  \cite{tasaki2020physics}. While magnetic order    breaks spin-rotation as well  as time-reversal symmetry,   chiral spin states (CSSs) \cite{wen1989chiral} epitomize the possibility of breaking   the latter while keeping  a vanishing expectation value for the spin operator.  For SU(2)-invariant systems, the scalar spin  chirality $\langle \mb S_i\cdot (\mb S_j\times \mb S_k)\rangle $ for three lattice sites  defines  an order parameter for CSSs.  This large class of states includes, in particular, topologically nontrivial chiral spin liquids with anyonic excitations \cite{wen1989chiral,Kalmeyer1987,Mudry1989,Laughlin1990,Greiter2009,Gong2014,Gong2015,Wietek2015,Cookmeyer2021}. For anisotropic exchange interactions,  the generalizations   of the scalar spin chirality are three-spin operators which  remain invariant under discrete spin rotations.  For instance, the operator $S_i^xS_j^yS_k^z$, invariant under global $\pi$ rotations about the $x$, $y$, and $z$ axes,  appears   in models with Kitaev-type  interactions  \cite{kitaev2006anyons,Yao2007,saket2010manipulating,Jianlong2019,Luo2022}.

Spontaneous     chirality and magnetization  are not necessarily competing orders, since they coexist in non-coplanar  phases of frustrated magnets \cite{Messio2011,Batista2016,Hayami2021}. On the other hand, the order parameter of a collinear magnetic phase such as the N\'eel state vanishes in a CSS and vice versa. While a CSS preserves spin-rotation symmetries,  a collinear  magnetic  state is usually invariant under  a combination of time reversal  and spatial or spin rotation that constitutes  a broken  symmetry in the  CSS.  As a result, in the absence of a coexistence region,    the Landau-Ginzburg-Wilson (LGW) paradigm dictates that  a generic phase transition from collinear magnetism to a CSS should be of  first order \cite{landau2013statistical,Imry1975}.  

Continuous phase transitions beyond  the LGW paradigm have been discussed in the context of deconfined quantum criticality (DQC) \cite{senthil2004deconfined,senthil2004quantum,levin2004deconfined,Wang2017}. The most  studied example is the continuous transition from the N\'eel state to a valence  bond solid (VBS) on the  square lattice \cite{Haldane1988,Read1989}. Such  an exotic transition can be described by effective field theories with a rich phenomenology that includes   order-parameter fractionalization, dualities,  and emergent higher symmetries  leading to non-compact gauge fields  and deconfined spinons at the quantum critical point. Unambiguous numerical demonstrations of DQC in lattice models \cite{Sandvik2007,Melko2008} have been hindered by logarithmic corrections to finite-size scaling, which   make it difficult to rule  out a  weakly first-order transition  \cite{NahumPRX,Shao2016,Ma2020}.  This challenge has motivated the study of Landau-forbidden transitions with analogies to DQC in one-dimensional (1D) models \cite{jiang2019ising,mudry2019quantum,Huang2019,Ogino2021,Roberts2021}, for which more controllable   analytical and numerical methods are available. In fact, it has been known for a while that the same operator that gives rise to  N\'eel order in the field theory for   anisotropic spin-1/2 chains can also generate spontaneous dimerization  \cite{Haldane1982}. The staggered magnetization and dimerization operators  can be combined into a  single order parameter, associated with the   SO(4) symmetry of the SU(2)$_1$ Wess-Zumino-Witten conformal field theory (CFT) at the Heisenberg point \cite{Nahum2015}, and the  continuous transition from the N\'eel  to the dimerized phase in  the frustrated  XYZ  chain has an emergent U(1) symmetry  \cite{jiang2019ising,mudry2019quantum}. 

In this work, we extend the set of Landau-forbidden phase transitions  by constructing lattice  models which exhibit a continuous transition from a CSS to a collinear magnetic phase. We start by unveiling a local mapping of the Hilbert space on a four-site unit  cell  that allows us to  represent the chirality and the magnetic order parameters as two  components of the same pseudospin.  As the main example of our construction, we consider a zigzag spin chain with  Kitaev interactions in addition to six-spin interactions that couple the   chiralities on   triangular plaquettes. For a particular choice of the parameters, our model reduces to the   one  proposed by Saket {\it et al.} \cite{saket2010manipulating} and becomes exactly solvable  in terms of Majorana fermions and a static $\mathbb Z_2$ gauge field. The six-spin interaction  stabilizes a chiral phase  as it   lifts the exponential ground state degeneracy of the  model of Ref. \cite{saket2010manipulating}.  In the regime of strong intercell Kitaev interactions   in the chain direction, we find a collinear magnetic phase analogous to the stripe phase of the Kitaev-Heisenberg   model on the triangular lattice  \cite{kimchi2014kitaev,Becker2015,Li2015,Kazuya2016,Maksimov2019}.  Thus,  our work also  fits in the context of recent studies of quasi-1D  extended Kitaev models aimed at offering insight into  two-dimensional phases \cite{LeHur2017,Agrapidis2018,agrapidis2019ground,Yang2020,Nocera2020,Sorensen2021,luo2021unveiling,Metavitsiadis2021}. We demonstrate the continuous transition between the CSS and the collinear magnetic phase using a combination of solvable effective Hamiltonians, bosonization of the low-energy theory, and numerical density matrix renormalization group (DMRG) simulations \cite{WhiteDMRG1992,white-dmrg-prb}. We show that the transition has an emergent U(1) symmetry and is described by a CFT with central charge $c=1$. We also analyze the transition from the point of view of a U(1) gauge theory with fermionic partons and   discuss  analogies with DQC in higher dimensions. 

The paper is organized as follows. In Sec.  \ref{sec:duality}, we present the  pseudospin  mapping and  apply it to the zigzag chain model. In Sec.  \ref{sec3}, we focus on the parameter regime in which the  model is exactly solvable by a Majorana fermion representation. In this case, we obtain a chiral ground state and  we classify the  excitations in terms of   fermionic modes and chirality domain walls. Section \ref{sec:transition} contains our analytical  results for the transition between chiral and magnetic phases.  We point out another exactly  solvable limit  of the model  and use  it as starting point for the   effective field theory, uncovering  some analogies with DQC. Our DMRG results which support the  prediction of a $c=1$ CFT at criticality  are presented in Sec. \ref{secDMRG}. Section \ref{sec2D} serves as an outlook, in which we offer some remarks about possible connections with  2D models  that harbor chiral spin liquid ground states. Finally, we summarize our findings  in Sec. \ref{secConclusion}.

\section{Chirality pseudospins and zigzag chain model\label{sec:duality}}

In this section, we  present a pseudospin mapping that will prove useful in studying chiral phases of spin systems with bond-dependent anisotropic exchange interactions, as in quantum compass models \cite{Nussinov2015}. We note in passing that a  remarkable duality between  the scalar spin chirality and the staggered-dimer order parameter  has been applied to interpret  the   chiral phase of the isotropic two-leg ladder with four-spin interactions  \cite{Hikihara2003,Lauchli2003,Gritsev2004}.

\begin{figure}[t]
\centering{}\includegraphics[width=\columnwidth]{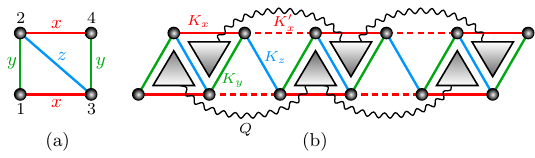}
\caption{Four-site unit cell and zigzag chain model. (a) The anisotropic chirality pseudospins on the  triangular plaquettes are defined according to the bond labels $x$, $y$, and $z$ as written in Eq. (\ref{taux}).   (b) In the zigzag chain,  spins are coupled by Kitaev   interactions. In addition, there is a six-spin interaction that couples the  chiralities  of  triangles with the same orientation (up-pointing or down-pointing) on  nearest-neighbor  unit cells. }
\label{fig1}
\end{figure}

Consider  Pauli spin operators  $\sigma_n^a$, with $a=x,y,z$, defined on four sites, $n=1,\dots,4$, represented as a square in Fig. \ref{fig1}(a). We choose the diagonal bond between sites $n=2$  and $n=3$ to divide the square into two triangles, and label the bonds as $x$, $y$, and $z$ so that each triangle contains one bond of each type. We then define the anisotropic spin chiralities on the triangles as the three-spin operators\be
\tau_1^x=\sigma_2^x\sigma_3^y\sigma_1^z,\qquad  \qquad \tau_2^x=\sigma_3^x\sigma_2^y\sigma_4^z.\label{taux}
\ee
The spin indices in $\sigma_i^a\sigma_j^b\sigma_k^c$ obey the mnemonic rule that, in the triangle formed by sites $(i,j,k)$, site $i$ corresponds to the vertex opposite to an $a$ bond, site $j$ is  opposite to a $b$ bond, and site $k$ is  opposite to a $c$ bond. We define the $z$ components of the pseudospins as \be
\tau_1^z=\sigma_1^x,\qquad \qquad  \tau_2^z=\sigma_2^x.\label{tauz}
\ee
The operators in Eqs. (\ref{taux}) and (\ref{tauz})  square to the identity   and  obey $[\tau_1^a,\tau_2^b]=0$, $\{\tau_1^x,\tau_1^z\}=\{\tau_2^x,\tau_2^z\}=0$.  A duality transformation that exchanges the chiralities with one-spin operators can be obtained by applying the unitary $U=e^{-i\pi (\tau_1^y+\tau_2^y)/4}$, where $\tau_1^y=i\tau_1^x\tau_1^z=-\sigma_2^x\sigma_3^y\sigma_1^y$ and  $\tau_2^y=i\tau_2^x\tau_2^z=\sigma_3^x\sigma_2^z\sigma_4^z$. 

To complete the mapping, we define another pair of Pauli operators which   commute with $\boldsymbol \tau_1$ and  $\boldsymbol \tau_2$. The first $\rho$ pseudospin  is defined by \bea
\rho_1^x=\sigma_2^x\sigma_3^y,\qquad \rho^y_1=\sigma_1^x\sigma_2^x\sigma_3^z,  \qquad \rho_1^z=\sigma_1^x\sigma_3^x, \label{rhoz1}
\eea
and the second one by
\bea
\rho_2^x=\sigma_2^x\sigma_4^y,\qquad  \rho^y_2=\sigma_4^z,\qquad \rho_2^z=\sigma_2^x\sigma_4^x.\label{rhoz2}
\eea
In both cases the $x$ and $z$ components  are time-reversal-invariant  two-spin operators, whereas the $y$ components   are time-reversal odd.   In contrast,  all components of $\boldsymbol \tau_1$ and  $\boldsymbol \tau_2$ are time-reversal odd.   Note that we do not refer to the three-spin operator $\rho_1^y$ as a spin chirality because it is not invariant under  $\pi$ rotations about the $x$ or $y$ axes. 

In order to illustrate the spin-chirality duality, we  consider a spin model defined on a zigzag chain,
given by the Hamiltonian \be
H = H_K+ H_Q,\label{HKQ}
\ee
where
\be
H_K= \sum_{a = x,y,z}K_a\sum_{\langle i,j\rangle_a} \sigma_i^a \sigma_j^a + K_x^\prime \sum_{\langle i,j\rangle_{x'}}\sigma^x_i\sigma^x_j 
\label{zigzaghamiltonian}
\ee
contains the bond-dependent Kitaev interactions. Here   $y$ and $z$ bonds couple sites on different  legs of the zigzag chain  and   $x$ bonds lie in the chain direction. We distinguish between two types of $x$ bonds. The bonds  with Kitaev coupling $K_x$ lie within a unit cell and are represented by solid red lines in Fig. \ref{fig1}(b). The $x$ bonds between sites in neighboring unit  cells have coupling $K_x^\prime$ and are represented by dashed red lines.  We focus on the regime of antiferromagnetic Kitaev couplings, $K_a,K_x^\prime\geq 0$. The zigzag chain with $K_x=K_x'=K_y=K_z$ can be viewed as a strip of the   Kitaev model on the triangular lattice, whose ground state has been controversial \cite{Becker2015,Li2015,Kazuya2016,Maksimov2019}. Besides the Kitaev interactions, we also add terms coupling the chiralities in different unit cells. The six-spin interaction preserves time-reversal symmetry and can be written in terms of the pseudospins in Eq. (\ref{taux}) as \be
H_Q=-Q\sum_{r}(\tau^x_{1,r}\tau^x_{1,r+1}+\tau^x_{2,r}\tau^x_{2,r+1}),\label{HQ}
\ee
where $\boldsymbol \tau_{1,r}$ and $\boldsymbol \tau_{2,r}$ denote  the pseudospins for the unit cell at position $r$. This interaction is analogous to the coupling between the scalar spin chiralities on plaquettes of the square lattice   proposed  in Ref. \cite{wen1989chiral}. Physically, this type of multi-spin interaction can be associated with orbital currents in chiral magnets \cite{Bulaevskii2008,Grytsiuk2020}. We focus on $Q\geq 0$,  favoring uniform chirality.  The reason for the particular choice of coupling only triangles with the same orientation, see Fig. \ref{fig1}(b), will become clear in Sec. \ref{sec:transition}.

The relevant symmetries of the model are:  time reversal $\mc{T}:  i \mapsto -i$, $\boldsymbol{\sigma}_{n,r} \mapsto - \boldsymbol{\sigma}_{n,r}$ for $n=1,\dots,4$;  a $C_2$ rotation symmetry about the center of a $z$ bond $\mc{R}: \boldsymbol{\sigma}_{1,r} \leftrightarrow \boldsymbol{\sigma}_{4,-r},\boldsymbol{\sigma}_{2,r} \leftrightarrow \boldsymbol{\sigma}_{3,-r} $; 
and two discrete  spin rotation symmetries $\mc{K}_1$ and $\mc K_2$ generated  by $U_1 = \prod_r \tau^x_{1,r}$ and $U_2 = \prod_r \tau^x_{2,r}$, respectively,  which can be understood as $\pi$ rotations about   a different spin axis for each sublattice \cite{Chaloupka2015}. Their combined action is known as the Klein symmetry
  because  it is isomorphic to the Klein group, $\mc K = \mc K_1 \times \mc K_2 \simeq \mathbb Z_2 \times \mathbb Z_2$  \cite{kimchi2014kitaev}. The Klein symmetry is   found in the Kitaev model on several  lattices that obey a certain geometrical condition, including the triangular lattice, but is explicitly broken by more general interactions such as Heisenberg exchange. The product $U_1U_2=\prod_{r}\prod_{n=1}^4\sigma_{n,r}^z$ generates the usual global $\pi$  rotation about the $z$ spin axis. 
   
In terms of the pseudospins, the Hamiltonian becomes\bea
H&=&\sum_{r}\bigg[\sum_{l=1,2}\left(K_x \rho_{l,r}^z+K_x'\rho^z_{l,r}\tau^z_{l ,r}\tau^z_{l,r+1}-Q\tau^x_{l,r}\tau^x_{l,r+1}\right) \nonumber\\
&&+K_y(\rho_{1,r}^x\rho_{2,r}^x-\rho_{1,r}^y\rho_{2,r}^y\tau^x_{1,r}\tau^x_{2,r})\nonumber\\
&&+K_z(\rho^x_{1,r}\rho^y_{2,r}\tau^x_{2,r}+\rho^y_{2,r} \rho^x_{1,r+1} \tau^x_{1,r+1} ) \bigg].\label{newH}
\eea   
This Hamiltonian is equivalent to a two-leg ladder with $l=1,2$ playing the role of a leg index.  There are two  pseudospins 1/2, namely $\boldsymbol \tau$ and $\boldsymbol \rho$, on each effective site specified by $(l,r)$. In the following sections, we  will  analyze  special limits of the model to establish the existence of a continuous phase  transition from a CSS in which $\langle \tau^x_{l, r}\rangle \neq 0$ to a  magnetic phase in which $\langle \tau^z_{l, r}\rangle \neq 0$.

\section{Chiral Spin State in the exactly solvable model\label{sec3}}   

In this section, we discuss the exact  solution of the model with $K_x^\prime=0$. In this case, the local operators $\tau_{l,r}^x$ commute with the Hamiltonian in Eq. (\ref{newH}), generating an extensive number of conserved quantities. For $Q>0$, the ground state has the same eigenvalue for all $\tau_{l,r}^x$, corresponding to a uniform spin chirality that   spontaneously breaks the $\mc T$ symmetry and preserves the $C_2$ rotation and Klein symmetries. Setting $\tau_{l,r}^x=1$, we obtain the effective Hamiltonian for the remaining pseudospins:\bea
H_{\rho}&=&\sum_{r}\bigg[K_x\sum_{l=1,2} \rho_{l,r}^z +K_y(\rho_{1,r}^x\rho_{2,r}^x-\rho_{1,r}^y\rho_{2,r}^y)\nonumber\\
&&+K_z(\rho^x_{1,r}+ \rho^x_{1,r+1}  ) \rho^y_{2,r}-2Q\bigg]\label{Hrho1}.
\eea    
The model is now equivalent to a single  chain   with anisotropic exchange couplings in the $xy$ plane and an effective field in the $z$ direction. We apply  the Jordan-Wigner transformation: \bea
\rho_{l,r}^z&=&1-2d^\dagger_{l,r}d^{\phantom\dagger}_{l,r},\nonumber\\
\rho_{1,r}^+&=&d_{1,r}\prod_l\prod_{r'<r}(1-2d^\dagger_{l,r'}d^{\phantom\dagger}_{l,r'}),\\
\rho_{2,r}^+&=&d_{2,r}(1-2d^\dagger_{1,r}d^{\phantom\dagger}_{1,r})\prod_{r'<r}(1-2d^\dagger_{l,r'}d^{\phantom\dagger}_{l,r'}),
\nonumber
\eea
where   $d_{l,r}$ are complex spinless  fermions and $\rho^\pm_{l,r}=(\rho^x_{l,r}\pm i\rho^y_{l,r})/2$. The Hamiltonian becomes  \bea
H_{\rho}&=&\sum_{r}\bigg[-2K_x\sum_{l=1,2}d^\dagger_{l,r}d^{\phantom\dagger}_{l,r} +2K_y(d^\dagger_{1,r}d^\dagger_{2,r}+\text{h.c.})\nonumber\\
&&+iK_z(d^\dagger_{1,r}-d^{\phantom\dagger}_{1,r})(d^\dagger_{2,r}-d^{\phantom\dagger}_{2,r})\nonumber\\
&&-iK_z(d^\dagger_{1,r+1}+d^{\phantom\dagger}_{1,r+1})(d^\dagger_{2,r}+d^{\phantom\dagger}_{2,r}) \bigg]+\text{const}.\label{Hrho}
\eea
It is then clear that the model admits an exact  solution in terms of free fermions.

The solution for  $K_x^\prime=Q=0$ was discussed in Ref. \cite{saket2010manipulating} using a Majorana fermion representation.  The key observation is that  in this  case  the zigzag chain reduces to  a tricoordinated 1D lattice, called the   tetrahedral chain, and the model can be solved  by the same methods  as  Kitaev's honeycomb model \cite{kitaev2006anyons}.  Using  Kitaev's representation, we  write the original spin operator at site $j=(n,r)$ as $\sigma_{j}^a=ib^a_{j}c^{\phantom{a}}_{j}$, where $b^a_{j}$ and $c^{\phantom{a}}_{j}$ are Majorana fermions subjected to  the local constraint $b^x_{j}b^y_{j}b^z_{j}c_j^{\phantom{x}}=1$. The tetrahedral-chain  Hamiltonian can be written as\be
H_0=\lim_{K_x^\prime\to0}H_K= \sum_{a = x,y,z}K_a\sum_{\langle j,l\rangle_a} iu^a_{jl} c_jc_l,\label{Majorana}
\ee
where   $u_{jl}^a=-ib_j^ab^a_l$ are locally conserved $\mathbb Z_2$ gauge fields defined on   the  $\langle j,l\rangle_a$ bonds, satisfying $u^a_{jl}=-u^a_{lj}$ and $[u^a_{jl},H_0]=[u^a_{jl},u^b_{j'l'}]=0$. The  chirality operators  can be identified with the $\mathbb Z_2$   fluxes in the triangles:\bea
\tau^x_{1}=u^x_{13}u_{32}^zu_{21}^y,\qquad \tau^x_{2}=u^x_{42}u_{23}^zu_{34}^y.
\eea 
Fixing a gauge with  $u_{jl}^a=\pm1$ so that $\tau^x_{1,r}=\tau^x_{2,r}=1$ for all unit cells, we find that the remaining $c$-type Majorana fermions  move freely in the background of the static gauge field.  
 Importantly, the spectrum also contains excitations which correspond to changing the $\mathbb  Z_2$ flux configuration. Starting from the state with uniform chirality $\tau^x_{l,r}=1$  $\forall l,r $, we refer to an excitation with a single $\tau^x_{l,r}=-1$ as a type-$l$ vortex of the $\mathbb  Z_2$  gauge field. In the solvable model,  type-1 and type-2  vortices are localized in up-pointing and down-pointing triangles, respectively, and  they can be created by changing the sign of a single $u_{jl}^x$. Note, however, that physical states must respect a global fermion parity constraint \cite{Pedrocchi2011}.

For $Q=0$, the exactly solvable model suffers from  a ground state degeneracy that grows exponentially with system size \cite{saket2010manipulating}. The reason is that the energies of the exact eigenstates only depend on the total $\mathbb Z_2$ flux in each unit cell, {\it  i.e.}, on  the product  $\tau_{1,r}^x\tau_{2,r}^x$, rather  than    the individual chiralities $\tau_{l,r}^x$. This degeneracy can be associated with a local symmetry as follows. For $Q=K_x^\prime=0$,   flipping the sign of both chiralities in   unit cell $r_0$ in Eq. (\ref{newH}) only affects the $K_z$ term that couples   $d_{1,r_0}$    to $d_{2,r_0}$ and $d_{2,r_0-1}$ in the quadratic Hamiltonian Eq. (\ref{Hrho}). The sign change can be removed by applying a gauge transformation $d_{1,r_0}\mapsto -d_{1,r_0}$. Thus, all eigenstates with  $\tau_{1,r_0}^x=\tau_{2,r_0}^x=-1$ have the same energy as the corresponding eigenstates  with   $\tau_{1,r_0}^x=\tau_{2,r_0}^x=1$.  In the Kitaev representation, this means that pairs of vortices located in the same unit cell cost zero energy for $Q=0$. The role of the chirality-chirality coupling  that we introduced in Eq. (\ref{HQ}) is therefore to impose an energy cost proportional to $  Q$ for   vortex pairs. For $Q>0$, there are only two    ground states  with uniform chirality   $\tau_{l,r}^x=\pm1$.

Let us now discuss the excitations in  the exactly solvable model. First, consider the    vortex-free sector  with  uniform chirality $\tau_{l,r}^x=1$. Fixing the set of $u_{jl}^a$ in a translationally invariant ansatz and taking a Fourier transform in Eq. (\ref{Majorana}),  we   cast the Hamiltonian in the form $H=i\sum_{k>0}C^\dagger_{k}\mc A(k)C^{\phantom\dagger}_{k}+\text{const.}$, where   $C_k^\dagger=(c^\dagger_{1,k},c^\dagger_{2,k},c^{ \dagger}_{3,k},c^{ \dagger}_{4,k})$ is a four-component   spinor in the  sublattice basis, with $c^\dagger_{n,k}=c^{\phantom\dagger}_{n,-k}$, and $\mc A(k)$ is a   matrix   obeying $\mc A^t(k)=-\mc A(-k)$. 
  The dispersion relations of the bands   can be calculated analytically for  arbitrary  values of the Kitaev couplings. The single-fermion energy gap is given by $\Delta_c= \sqrt{K_x^2+K_y^2+K_z^2-2K_z\sqrt{K_x^2+K_y^2}}$  and vanishes for $K_z^2= K_x^2+K_y^2$ \cite{saket2010manipulating}.   In the  fermionic representation, the gap closing involves  a change in    the $\mathbb  Z_2$ topological invariant $\nu=\text{sgn}\{\text{Pf}[\mc A(0)]\text{Pf}[\mc A(\pi)]\}$ 
 for  class D in  one dimension \cite{Chiu2016}.  The  phase for $K_z^2> K_x^2+K_y^2$ corresponds to the nontrivial  value  $\nu=-1$. 
 However, in the spin representation  this transition can be  associated with local order parameters. According to the Hamiltonian in  Eq. (\ref{Hrho1}), the limit of dominant $K_z$ corresponds to the $\boldsymbol \rho$ pseudospins ordering in the $xy$ plane. Importantly, nonzero expectation values of $\rho^x_{l,r}$ and $\rho^y_{l,r}$  break spin-rotation  symmetries.   Since time-reversal symmetry is already broken by the spontaneous  spin chirality, we obtain $\langle \sigma^z_{1,r}\rangle\sim \langle \tau^x_{1,r}\rangle  \langle \rho^x_{1,r}\rangle\neq 0$. Thus,  for $K_z^2> K_x^2+K_y^2$ we encounter a phase with spontaneous magnetization in the $z$ spin direction coexisting with the spin chirality. Since this phase breaks more symmetries than the pure chiral phase for $K_z^2< K_x^2+K_y^2$, this  is a conventional Ising transition and we do not  discuss it further. Note that  the   gap for   $c$  fermions closes at the critical point,  but   vortex excitations remain gapped across  this transition.
 
 \begin{figure}[t]
\centering{}\includegraphics[width=\columnwidth]{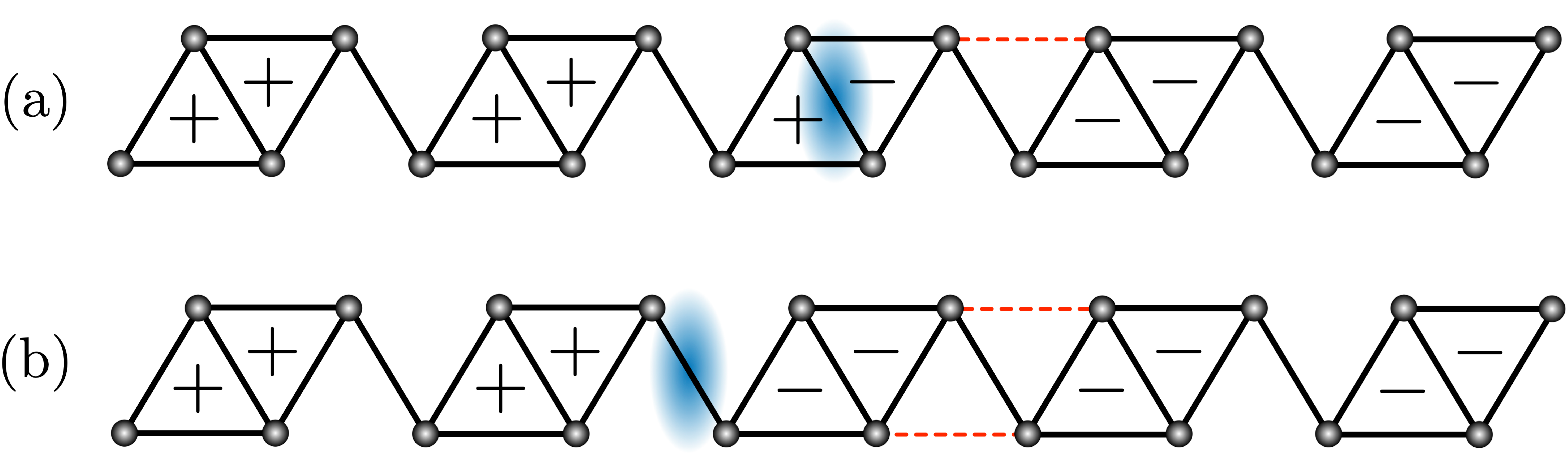}
\caption{Two types of chirality domain walls.  The signs indicate the chirality $\tau_{l,r}^x$ for each triangle.  (a) In the nonintegrable model with   $K_x^\prime\ll K_a,Q$, an intracell domain wall can hop one  unit cell to the right by applying the $K_x^\prime$ interaction on the $x'$ bond indicated by the dashed   line. (b) An intercell domain wall moves two unit cells to the right  when the perturbation acts   on the two bonds indicated by the dashed lines. }
\label{fig2}
\end{figure}

We now turn to  vortex excitations. In this  case, we compute  the energies  numerically by diagonalizing the Hamiltonian on a finite chain with open  boundary conditions because the localized vortex breaks translational invariance.   The energy of a single vortex behaves as $\Delta_{v}=\Delta^{(0)}_{v} +2Q$, where $\Delta^{(0)}_{v}$ denotes the vortex gap for $Q=0$.  In particular, for  $K_x=K_y=K_z$ we obtain $\Delta^{(0)}_{v} \approx 0.29K_x$.   The vortex gap vanishes if any of the Kitaev couplings $K_a$ approach zero since  the broken   bonds make the $\mathbb Z_2$  fluxes ill defined. For $K_y=K_z\ll K_x$,  we  find  $\Delta^{(0)}_{v}\approx  K_y^2/(2K_x)$.  

In the pseudospin representation, it may be more   convenient to  picture   the elementary excitations in  the flux sector   as domain walls between regions of opposite chiralities. We distinguish between two types of domain walls as illustrated in Fig. \ref{fig2}. We call an intracell domain wall the situation in which the two domains are separated by a unit cell $r_0$ with $\tau^x_{1,r_0}\tau^x_{2,r_0}=-1$; see Fig. \ref{fig2}(a).  By contrast, in an intercell domain wall the chirality switches between two adjacent unit cells so that   $  \tau^x_{1,r}\tau^x_{2,r}=1$ for all unit cells in the vicinity of the domain wall; see Fig. \ref{fig2}(b).  Starting from a ground state with uniform chirality, we create an  intercell domain wall at position $r_0$  by  applying the string operator $V_{r_0}= \prod_{r< r_0} \tau_{1,r}^z\tau_{2,r}^z$, with  energy cost $2Q$. An intracell domain wall with energy $\Delta_{v}$ is created by $V_{r_0}\tau_{1,r_0}^z$ or $V_{r_0}\tau_{2,r_0}^z$. A $\mathbb Z_2$ charge for the domain walls can be defined as the eigenvalue of $\prod_{r}\tau^x_{1,r}\tau^x_{2,r}=U_1U_2$; recall that this  is the generator of $\pi $ rotations about the $z$ axis. The intracell domain wall  is the one which  transforms nontrivially under this $\mathbb Z_2$ symmetry.

\begin{figure}[t]
\centering{}\includegraphics[width=.95\columnwidth]{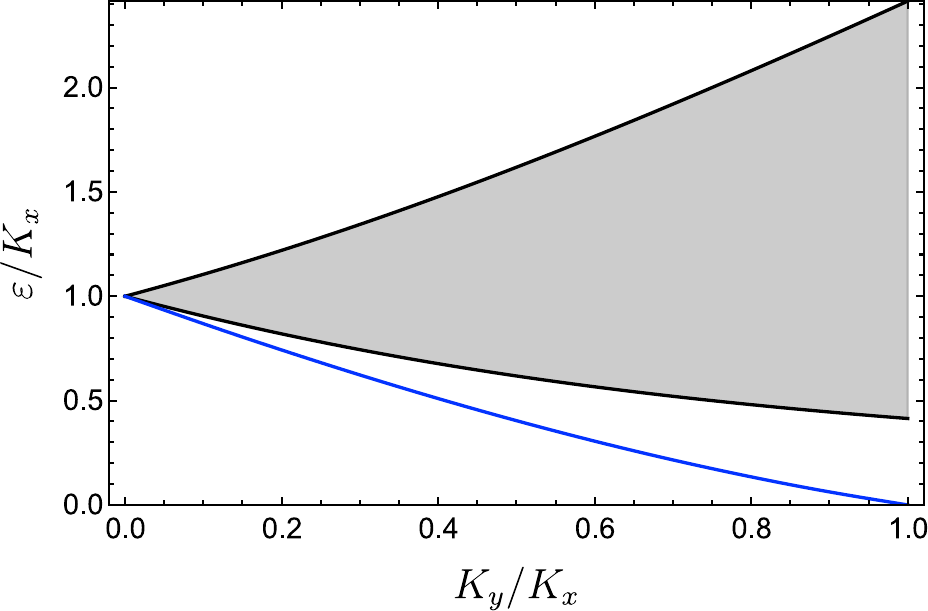}
\caption{Fermionic spectrum calculated from the Hamiltonian in Eq. (\ref{Majorana}) for an open chain containing  a single   intracell domain wall. Here we set $K_z=K_y$.  The shaded region  represents the continuum of extended states. The   blue line represents the  energy of the bound state.  }
\label{fig3}
\end{figure}

The creation  of   chirality domain walls affects the fermionic spectrum. Here   we restrict the parameters to the domain  $K_y=K_z\leq K_x$. This condition    puts   the system in  the pure chiral  phase with $\nu=1$.  We observe numerically that in the presence of   an intercell domain wall the fermionic spectrum only comprises  a continuum of extended states, as in the vortex-free case. On the other hand, the creation of  an intracell domain wall gives rise to a midgap state   in which a   $c$   fermion is bound near the unit cell with $  \tau^x_{1,r}\tau^x_{2,r}=-1$. The energy of the bound state is not pinned at zero, but varies with the ratio $K_y/K_x$ as shown in Fig. \ref{fig3}. In particular, the bound state has zero energy for $K_x=K_y=K_z$. We expect  the transition from the chiral phase to a non-chiral magnetic phase  to be associated with  the condensation of domain walls which   carry the magnetization  degree of freedom in the form of a bound state of  a $c$-type matter  fermion and a $b$-type flux fermion. This transition will be discussed  in the next section.

 
\section{Transition to the magnetic state \label{sec:transition}}

For $K_x'>0$, the local chirality operators are no longer conserved because the operator $\tau^z_{l,r}\tau^z_{l,r+1}$ in Eq. (\ref{newH}) flips the chirality of two  triangles with the same orientation in neighboring unit cells. Nevertheless,  the  parity of the number of type-$l$ vortices, encoded in the eigenvalues of  $U_l=\prod_r \tau^x_{l,r}$, are still good quantum numbers. Clearly, the existence of two separately conserved parities   is associated with   the Klein symmetry. In the    regime    $K_x'\ll K_a,Q$, we can treat the $K_x'$ term   as a perturbation to the solvable model discussed in Sec. \ref{sec3} and focus on   a subspace with a fixed number of vortices or domain walls. Figure \ref{fig2} shows the processes that  generate an  effective hopping amplitude for  the domain walls. While the intracell domain wall moves at first order in $K_x^\prime$, the intercell domain wall only moves at order $(K_x')^2$. As a result, both types  acquire a dispersion, but the intracell domain wall has a larger bandwidth. We then expect that, as we increase $K_x^\prime$ in the regime $Q\gtrsim \Delta_v^{(0)}$, the gap for intracell domain walls will eventually close first, driving a phase transition. Below the critical value of $K_x'$, the mobile domain walls remain gapped and  the system is in a CSS characterized by $\langle \tau^x_{l,r}\rangle \neq 0$.  In the Kitaev representation,  the gauge variables $u_{jl}^y$ and $u_{jl}^z$ still commute with the Hamiltonian, but $u_{jl}^x$ become fluctuating,  and the chiral phase corresponds to  $\langle u_{jl}^x\rangle\neq 0$.   

To examine the phase transition, let us consider the limit of highly  anisotropic Kitaev interactions. For $K_y,K_z\to0$, the $\boldsymbol \rho$ pseudospins are fully polarized with $  \rho^z_{l,r} = -1$ in the ground state. This  condition   imposes strong antiferromagnetic correlations in  the $x$ spin direction  between two spins  on  the  same leg and  in the  same unit cell; see Eqs. (\ref{rhoz1}) and (\ref{rhoz2}).   In the regime $K_y,K_z\ll K_x$, we can  safely assume that the $\boldsymbol \rho$ pseudospins are gapped out. We derive an effective Hamiltonian in the low-energy subspace   by applying perturbation theory to second order in $K_y$ and $K_z$. We obtain \bea
H_{\rm eff}&=&-\sum_{r} \sum_{l=1,2}\left(  K_x' \tau^z_{l ,r}\tau^z_{l,r+1}+Q\tau^x_{l,r}\tau^x_{l,r+1}\right) \nonumber\\
&&-K_\perp\sum_r \tau^x_{1,r}\tau^x_{2,r}+\text{const.} ,\label{Heff}
\eea   
where $K_\perp\approx K_y^2/(2K_x)$. This Hamiltonian describes a two-leg ladder with weakly coupled XY  chains.   Note that the interchain coupling   only involves the $x$ components of the pseudospins. The interactions $ \tau^a_{1,r}\tau^a_{2,r}$ with $a=y,z$ are forbidden by the Klein symmetry. 

 \begin{figure}[t]
\centering{}\includegraphics[width=.95\columnwidth]{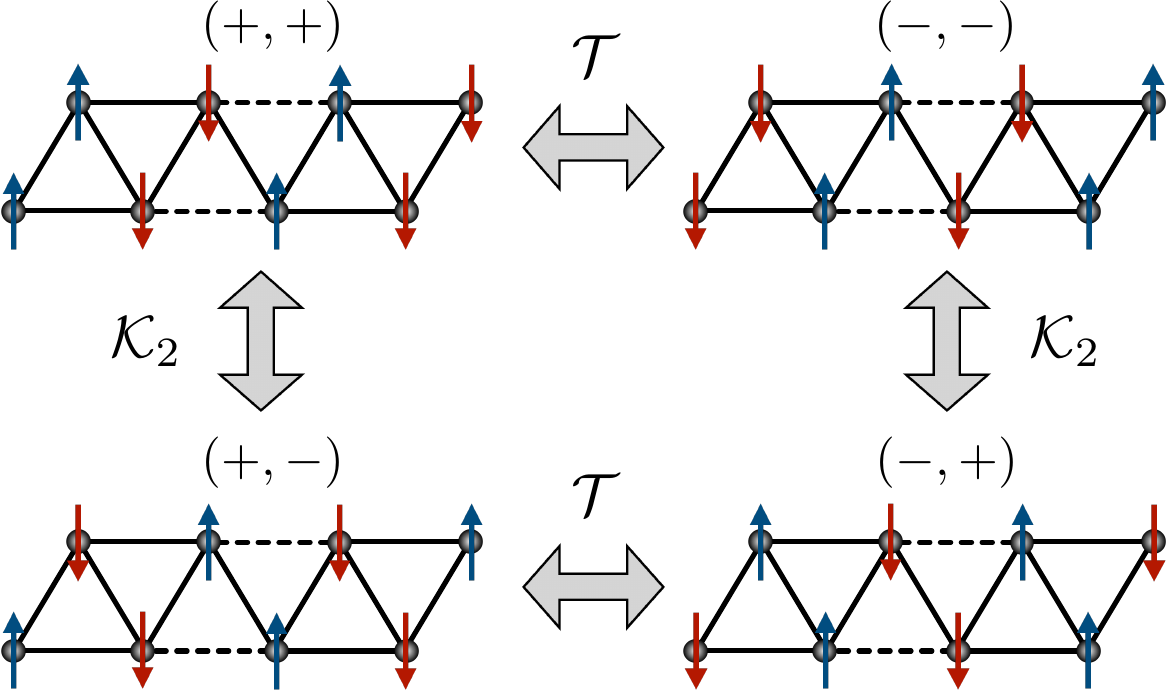}
\caption{Four    ground states in the magnetic phase. The up and down arrows represent the magnetization in the $x$ spin direction.  The symmetries that connect the different states are also indicated.  }
\label{fig4}
\end{figure}

This analysis reveals that the model with $K_y=K_z=0$ and $K_x,K_x^\prime,Q>0$ is also exactly solvable. Setting $K_\perp=0$, we see that  the Hamiltonian reduces to two identical and decoupled  XY chains, even though spins on different legs of the original zigzag chain are coupled by the six-spin interaction. The exact excitation spectrum can be calculated by performing a Jordan-Wigner transformation. The critical point  occurs at $K_x^\prime=Q$. The chirality vanishes continuously as we approach the critical point from $K_x'<Q$. For $K_x'>Q$, the system enters another ordered phase characterized by $\langle \tau^z_{l,r}\rangle\neq 0$.  Since $\rho_{l,r}^z=-1$, this implies    $\langle \sigma^x_{1,r}\rangle=-\langle \sigma^x_{3,r}\rangle\neq 0$ and $\langle \sigma^x_{2,r}\rangle=-\langle \sigma^x_{4,r}\rangle\neq 0$. Thus, we find a collinear magnetic phase with spontaneous magnetization in the $x$ spin  direction. There are four ground states, labeled by $(\text{sgn}\langle \sigma^x_{1,r}\rangle,\text{sgn}\langle \sigma^x_{2,r}\rangle )$ and represented in Fig. \ref{fig4}. The magnetic order in the zigzag  chain  is analogous  to the stripe phase of the antiferromagnetic Kitaev model on the triangular lattice as obtained in Ref. \cite{Maksimov2019}.  The pair of states $(+,+)$ and $(-,-)$ are  conjugated by $\mc T$ symmetry, and likewise for the pair $(+,-)$ and $(-,+)$. The $(+,+)$ and $(-,-)$ states break the $\mc R$ spatial rotation symmetry, but preserve $  \mc R\mc T$. The degeneracy between   $(+,+)$ and $(+,-)$  is protected by the  $\mc K_2$ symmetry; both $\mc K_1$ and $\mc K_2$   are spontaneously broken in this phase. All  four  ground states are invariant under the combination of time reversal and global $\pi$ rotation about the $z$ spin axis,  a symmetry broken in the CSS.  Remarkably, the $\mc T$ symmetry is broken on both sides of the transition, but restored at criticality.  In the magnetic phase, the elementary excitations are kinks or domain wall in the staggered magnetization. The magnetization domain walls  are mobile for any  $Q>0$ and condense when we approach  the phase transition from $K_x'>Q$.

We   determine the universality class of the  transition  by taking the continuum   limit in the effective Hamiltonian. Starting from $K_\perp=0$, we    bosonize the pseudospins in the XY chains following   the  standard procedure   \cite{giamarchi2003quantum} and then add the interchain coupling $K_\perp$ as a perturbation. We obtain the  Hamiltonian density 
\bea
\mc H_{\rm bos}&=&\sum_l \left[\frac{v\kappa}{2}(\partial_x\theta_l)^2+\frac{v}{2\kappa}(\partial_x\phi_l)^2+\lambda \cos(\sqrt{4\pi}\theta_l)\right]\nonumber\\
&&-\lambda_\perp \cos(\sqrt\pi\theta_1) \cos(\sqrt\pi\theta_2),\label{Hbos}
\eea
where the bosonic fields satisfy $[\phi_l(x),\partial_{x'}\theta_{l'}(x')]=i\delta_{ll'}\delta{(x-x')}$, $v$ is the velocity of the pseudospin excitations,  $\kappa$ is the Luttinger parameter, and $\lambda,\lambda_\perp>0$ are the coupling constants of the relevant operators with scaling dimensions $1/\kappa$ and $1/(2\kappa)$, respectively. In the vicinity of the transition, we have $v\sim K_x'$,  $\lambda \sim K_x'-Q$,   $\lambda_\perp \sim K_\perp$, and $|\kappa-1|\sim (K_\perp/K_x')^2$. The bosonized pseudospin operators can be written as  $\tau^x_{l,r}\sim \cos(\sqrt\pi\theta_l)$ and $\tau^z_{l,r}\sim \sin(\sqrt\pi\theta_l)$. 

For $\lambda_\perp=0$, the bosonic Hamiltonian reduces to two decoupled sine-Gordon models. The critical point at $\lambda=0$ contains two massless bosons and is described by a CFT with central charge $c=2$. The continuous symmetry $\theta_l\mapsto \theta_l +\text{const.}$ of the effective field theory at the critical point can be traced back to  the   lattice model. For  $K_y=K_z=0$ and $Q=K_x'>0$, the   operators  $Y_l=\sum_r\tau^y_{l,r}$ with $l=1,2$ become conserved charges in the sector with $\rho^z_{l,r}=-1$. Away from the critical point, the relevant cosine perturbation flows to strong coupling under the renormalization group.  The chiral phase corresponds  to $\lambda<0$, with $\lambda\to-\infty $ at low energies pinning the $\theta$ field at  $\theta_l=0$ or  $\theta_l=\sqrt \pi$.  For $\lambda>0$, the flow to $\lambda\to\infty$ pins the bosonic fields at $\theta_l= \sqrt \pi/2$ or $\theta_l= 3\sqrt \pi/2$, corresponding to the magnetic phase. For $\lambda_\perp=0$, both phases have a fourfold degenerate ground  state. The extra ground state degeneracy of the chiral phase   is due to the  decoupling of the chiralities with  different $l$  for $K_\perp=0$ in Eq. (\ref{Heff}).

When we switch on  $\lambda_\perp >0$, the bosonic fields  on  different legs  are coupled by  a strongly relevant  operator.  According to the $c$ theorem \cite{Zamolodchikov1986}, the central charge of the CFT  must decrease.   Since the cosine operators commute with each other, we can  proceed with a semiclassical analysis  of the effective potential $V(\theta_1,\theta_2)=\lambda\sum_l \cos(\sqrt{4\pi}\theta_l)-\lambda_\perp \cos(\sqrt\pi\theta_1) \cos(\sqrt\pi\theta_2)$. For $\lambda_\perp>0$, the potential only becomes flat in the directions $\theta_2=\pm \theta_1$ (mod $2\sqrt \pi$) of the $(\theta_1,\theta_2)$ plane when $\lambda=\lambda_\perp/4>0$, as opposed to a completely flat potential when $\lambda=\lambda_\perp=0$. This means that  the interchain coupling leaves out only one gapless boson  at the transition, either $\theta_+=(\theta_1+\theta_2)/\sqrt2$  or $\theta_-=(\theta_1-\theta_2)/\sqrt2$. As a  consequence, the generic  transition  for $\lambda_\perp>0$ has central charge $c=1$, associated with a  single emergent U(1) symmetry.  This  result is similar to the N\'eel-VBS transition in frustrated   spin chains \cite{jiang2019ising,mudry2019quantum}. At criticality, the correlation functions for both order parameters decay as power laws with the same exponent. Note that the transition at a critical value of $\lambda>0$  enlarges the region occupied by the chiral phase in comparison with  the result for $\lambda_\perp=0$.

We can  understand the  ground  state degeneracy by pinning  the bosonic fields in the presence  of the  $\lambda_\perp$ interaction.  Assuming that   $\lambda_\perp$   gaps out $\theta_-$, we fix $\theta_2=\theta_1$. In the chiral phase, this condition implies $\theta_1=\theta_2=0,\sqrt\pi$;  the two choices correspond to the  ground states with either sign of $\langle \tau^x_{1,r}\rangle=\langle \tau^x_{2,r}\rangle\neq0$. If we assume instead that $\lambda_\perp$ gaps out $\theta_+$ and fix $\theta_2=-\theta_1$, we find precisely the same  expectation values for the local physical operators.  Thus, the ground state of the chiral phase is twofold degenerate for $\lambda_\perp>0$. On  the other hand,  the choice of gapping out $\theta_+$ or $\theta_-$ affects the expectation value of the magnetization when we pin the bosonic fields at $\theta_1=\pm\theta_2=   \sqrt\pi/2, 3 \sqrt\pi/2$. In this case, there are still four possibilities labeled by  the  signs  of   $\langle \tau^z_{1,r}\rangle $ and $\langle \tau^z_{2,r}\rangle $. Provided that the Hamiltonian preserves the Klein symmetry, the ground state of the magnetic phase remains fourfold degenerate. In fact, the effective field theory   allows us to analyze the effects of breaking the Klein symmetry, which in the bosonic representation acts as $\mc K_l:\phi_l\mapsto -\phi_l, \theta_l \mapsto -\theta_l $. Adding the perturbation $\lambda_\perp^\prime \sin(\sqrt\pi\theta_1)\sin(\sqrt\pi\theta_2)$ to the Hamiltonian density in Eq. (\ref{Hbos}), we find that the total potential still pins $\theta_2=\pm\theta_1$ and leaves out one gapless boson with  $c=1$ at the transition. However, the ground state degeneracy of the magnetic phase is reduced to twofold, as the new interaction selects either $(+,+)$ and $(-,-)$ or $(+,-)$ and $(-,+)$,  depending on the sign of $\lambda_\perp^\prime$.

In the bosonic Hamiltonian  Eq. (\ref{Hbos}),  we dropped the  symmetry-allowed   cosine operators  such as $\cos(\sqrt{16\pi}\phi_l)$ because they are highly 
irrelevant for $\kappa\approx 1$. Vertex operators of the form $\exp(i m\sqrt{\pi}\phi_l)$ with $m\in\mathbb Z$ create or annihilate domain walls, which in the bosonic theory  correspond to kinks and antikinks in  the field configuration, $\theta_l(x \to \infty)-\theta_l(x \to -\infty) = \pm\sqrt \pi$.  In a  semiclassical picture for the chiral phase, to go from the ground state with $\theta_l=0$ to $\theta_l=\pm \sqrt \pi$, the bosonic fields have to go through $\theta_l=\pm \sqrt\pi/2$, which can be interpreted  as  the magnetization   $\langle \tau^z_{l,r} \rangle\sim\langle \sin (\sqrt\pi\theta_l)\rangle$ residing at the topological defect of the CSS. The same argument can be used  to see how domain walls in the magnetic phase must carry  spin chirality.  Near the transition,  the processes that  change the number of domain walls in either picture become irrelevant.

Similar phenomenology is generally found in effective field theories for DQC  \cite{senthil2004deconfined,senthil2004quantum,levin2004deconfined,Wang2017}: 
topological defects in one phase nucleate  the  order parameter of the other phase. If the   order parameters are
written in terms of   fractionalized excitations, the resulting constraints lead to an emergent gauge field on which the topological defects are charged. 
One can then understand both phases as distinct confined regimes, merging in a critical region corresponds to a gapless phase in the gauge theory.

In our model, the domain wall description can be obtained by a refermionization of the bosonic fields, defining the chiral fermions  $\psi_{R/L,l} \sim \exp\left[-i\sqrt \pi (2 \phi_l \mp  \theta_l /2)\right]$. The physical spin operators are then  given by fermion bilinears.  In terms of the two-component spinors  $\Psi^\dagger_l = (\psi^\dagger_{L,l}, \psi^\dagger_{R,l})$, we have $\tau^x_{l,r} \sim \Psi^\dagger_l \sigma^x \Psi^{\phantom\dagger}_l$ and
$\tau^z_{l,r} \sim \Psi^\dagger_l \sigma^y \Psi^{\phantom\dagger}_l$, with $\sigma^a$ the Pauli matrices acting in the internal space. The effective Hamiltonian includes   density-density  interactions which   arise from the cosine operators as well as quadratic terms  in Eq. (\ref{Hbos}). The emergent symmetry at the critical point  is manifested as Noether charges of the fermions, preventing pairing terms from appearing in the Hamiltonian. A mean-field decoupling of the quartic interactions generates mass terms for the chiral fermions in the ordered phases.   Solitonic configurations in the mass terms support  fermion bound states via  the Jackiw-Rebbi mechanism \cite{jackiw1976solitons}, confirming the  previous interpretation in terms of  domain walls. Note that this mechanism applies to smooth domain walls in the low-energy theory for the transition, as opposed to the sharp domain walls deep in the chiral phase discussed in Sec. \ref{sec3}. For a smooth domain wall, a zero-energy bound state is formed even if  the phase is topologically trivial \cite{robinson2019nontopological}. 

To describe the transition in the  fermionic picture,  we start from the assumption of an emergent U(1)$\times$U(1) symmetry, which can then be gauged.  The coupling to a  U(1) gauge field can be obtained by noticing  that the representation of the physical operators has a   gauge redundancy,  $\Psi_l(x)\mapsto e^{ie_l \Lambda_l(x)}\Psi_l(x)$,  where $e_l$ play the role of gauge charges. We then impose a   constraint on the   fermion densities    $\Psi^\dagger_l   \Psi^{\phantom\dagger}_l\sim \partial_x \theta_l$,  as usual  in parton constructions \cite{Wenbook}. The resulting gauge-invariant lagrangian   has the form \be
\mathcal{L} =  \sum_{l=1,2} \bar{ \Psi}_l i \gamma^\mu(\partial_\mu -ie_l a_\mu)  \Psi_l +\frac{1}{4g^2}(f_{\mu\nu})^2 +\cdots,\label{fermiongauge}
\ee
where we introduced the Maxwell tensor $f_{\mu\nu} = \partial_\mu a_\nu-\partial_\nu a_\mu$, the parameter $g$ in the Maxwell term controls the fluctuations of the gauge field,  and we omitted    quartic terms associated  with short-range interactions.  The  terms highlighted in Eq. (\ref{fermiongauge}) comprise the $N_f = 2$ Schwinger model in $1+1$ dimensions, known to reduce to a single massless boson at low energies \cite{hosotani1997gauge, kim1999theory, sheng2008strong}.  The gauge charges can  be  chosen arbitrarily as  $e_l=\pm1$. The relative sign between $e_1$ and $e_2$ selects symmetric or antisymmetric  modes with  respect  to exchanging  the leg  index, and is analogous  to pinning either $\theta_+$ or $\theta_-$   in the bosonic  theory.  The critical  point corresponds to fine tuning the quartic interactions  so that the  bosonic mode remains gapless, as described by Eq. (\ref{Hbos}) after we fix $\theta_2=\pm \theta_1$.  Once again, we come to the conclusion that the transition between the chiral and magnetic phases is described by a $c=1$ CFT.  At the fixed point, the two chiral sectors of the gapless boson   decouple, and the CFT has an enlarged U(1)$\times$U(1) symmetry \cite{Affleck1985}.

\section{Numerical Results\label{secDMRG}}

In this section, we present our DMRG results for the phase transition between the CSS and the collinear magnetic state. To investigate the phases and the nature of the phase transition, we   consider the zigzag chain with six-spin interactions described by the Hamiltonian in Eq.  (\ref{HKQ}), equivalent to Eq. (\ref{newH}), and the two-leg XY ladder defined in Eq. (\ref{Heff}). In particular,  we    show results for the expectation values $\bra\tau^{x,z}_{l,r}\ket$, the susceptibility of the ground-state energy density, and the entanglement entropy (EE). 
 
 \begin{figure}[t]
\centering{}\includegraphics[width=\columnwidth]{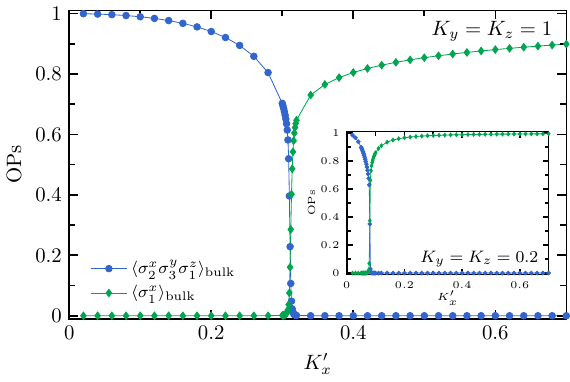}
\caption{Anisotropic spin chirality   and local magnetization   for the zigzag chain model, see Eqs. (\ref{HKQ})-(\ref{HQ}), as a function of $K_x'$ for $K_x=Q=1$ and $K_y=K_z=1$. The inset shows the same order parameters for $K_y=K_z=0.2$. }
\label{fig5}
\end{figure}

To compute the physical properties of interest, we have considered open chains with a maximum length of $L=400$. Keeping up to 400 states to represent the truncated DMRG blocks, we find that the largest truncation error acquired is smaller than $10^{-9}$ at the final sweep. As   discussed in Secs. \ref{sec3} and \ref{sec:transition}, both  chiral and magnetic  phases exhibit degenerate  ground states. Thus, to avoid linear combinations of the ground states in the numerical simulations, we have included weak and suitable perturbations that  couple to the order parameters at the chain edges and select one ground state for a given phase. These small local perturbations do not affect the bulk properties, probed by observables  computed  near the middle of the chain.

Let us first focus on the    zigzag chain model with six-spin interactions given by Eq.  (\ref{HKQ}). In Fig. \ref{fig5}, we show the bulk values for  the chiral  and magnetic order parameters as a  function of $ K'_x$ for $K_a=Q=1$, with $a=x,y,z$. While the CSS is characterized by a finite anisotropic spin chirality, the magnetically ordered phase displays an antiferromagnetic alignment along the $x$ spin direction (see Fig. \ref{fig4}). Note  that there  is  a single  phase transition at a critical value of $K_x'$. We have also  considered the regime $K_y=K_z<K_x$ and found that  the chiral phase becomes narrower as we decrease the ratio $K_y/K_x$, but the behavior is qualitatively the same as for $K_x=K_y=K_z$. No other transitions are observed as we vary $K_y/K_x$  for fixed $Q=K_x$; see the inset in Fig. \ref{fig5}.

\begin{figure}[t]
	\centering{}\includegraphics[width=\columnwidth]{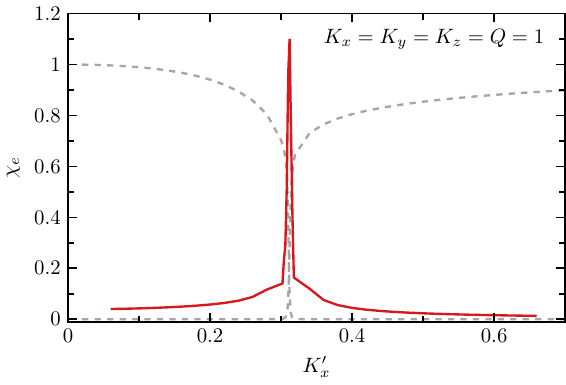}
	\caption{Susceptibility of the ground-state energy density as a function of $K_x'$ for $K_x=K_y=K_z=Q=1$. The dashed lines  represent the order parameters shown in the main plot of Fig. \ref{fig5}. The divergent behavior of $\chi_e$ at $K'_x\approx 0.312$ determines the critical point.}
	\label{fig6}
\end{figure}

To pinpoint the location of the phase transition, we have analyzed the energy susceptibility, defined as \be
\chi_e=-\frac{\partial^2 e_0}{\partial K_x'^2},
\ee
where $e_0$ is the ground-state energy per site.   In $d$ dimensions, the energy susceptibility diverges at the critical point as a power law with exponent  $\alpha=(2/\nu)-(d+z)$, where $\nu$ and $z$ are the correlation and the dynamical critical exponents   \cite{Albuquerque-scaling-susc,Sorensen2021}. In Fig. \ref{fig6}, we show $\chi_e$ as a function of $K_x'$ for $K_a=Q=1$. Note that $\chi_e$ exhibits a prominent peak, whose position in the $K_x'$ domain determines the critical point $K_x'=K'_{x,\text{crit}}$. For the set of    couplings shown in Fig. \ref{fig6}, we obtain $K'_{x,\text{crit}}\approx0.312$. To verify the accuracy of  the critical points extracted from $\chi_e$, we have also estimated $K'_{x,\text{crit}}$ from the analysis of the inflection point of the order parameters and the highest Schmidt eigenvalue. The latter was proposed in Ref. \cite{Sorensen2021}  as a   sensitive measure to  detect  phase transitions. Altogether, we found excellent   agreement among the estimates obtained from these distinct procedures.

We now turn   to the effective Hamiltonian in Eq. (\ref{Heff}),  valid   in the regime   $K_y,K_z\ll K_x$. This model  describes  two weakly coupled XY chains with interchain coupling along the $x$ direction. In comparison  with the original zigzag chain in Eq. (\ref{HKQ}), the dimension  of the local Hilbert  space in the  effective ladder model is reduced by a factor of 2, providing  a significant advantage for numerical simulations.  Since we observed the same qualitative behavior for the  original model with  $K_x=K_y=K_z$ as for small $K_{y,z}/K_x$, see Fig. \ref{fig5},  we expect   the effective XY ladder model in Eq. (\ref{Heff}) to capture the essential characteristics of the phase transition.   Carrying  out the same analysis as for the zigzag  chain, we again find only one transition for fixed $Q$ and different values of $K_x'$. In agreement with  the analysis in Sec. \ref{sec:transition}, the transition for $K_\perp>0$ shifts to larger values of $K_x'$  as compared to $K_x'=Q$  in the exactly solvable case $K_\perp=0$. 
Setting $Q=1$, we determined the critical point $K_\perp=K_{\perp,\text{crit}}$ for $K_x'=1.2$ and  $1.4$. The acquired values are $K_{\perp,\text{crit}}\approx0.187$ and $0.49$, respectively.

\begin{figure}[t]
	\centering{}\includegraphics[width=\columnwidth]{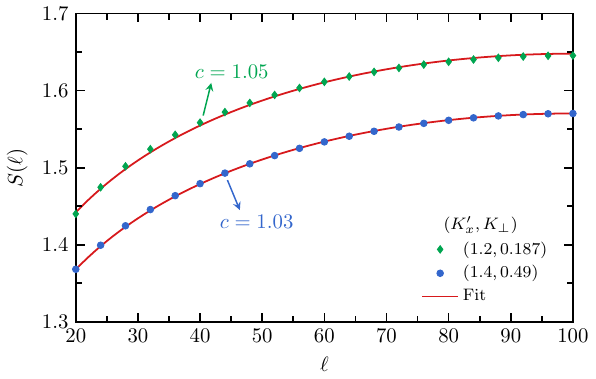}
	\caption{Entanglement entropy as a function of partition size $\ell$ for the critical two-leg XY ladder model, see Eq. (\ref{Heff}), with  $(K_x',K_\perp)=(1.2,0.187)$ and $(1.4,0.49)$. The symbols   represent  the DMRG results for  chain   length $L=200$ and open boundary conditions. The  solid red lines are fits to our numerical data using Eq. (\ref{entang}). The estimates for the central charge are indicated in the plot.} 
	\label{fig7}
\end{figure}

We  investigate the universality class of the transition by extracting the central charge from the EE. Consider a chain composed of two partitions $\mathcal{A}$ and $\mathcal{B}$ with $\ell$ and $L-\ell$ sites, respectively. The EE is then defined as $S(\ell)=-\mathrm{Tr}(\rho_\mathcal{A}\ln\rho_\mathcal{A} )$, 
where $\rho_\mathcal{A}$ is the reduced density matrix of   partition $\mathcal{A}$. For critical 1D systems, the asymptotic behavior of $S(\ell)$ predicted from CFT is given by \cite{Calabrese2004}
\be
S(\ell)=\frac{c}{3\eta}\ln\left[\frac{L}{\pi}\sin\left( \frac{\pi}{L}\ell\right)\right]+b,\label{entang}
\ee
where $c$ is the central charge, $b$ is a nonuniversal constant, and $\eta=1(2)$ for periodic (open) chains.  In Fig. \ref{fig7}, we show the EE as a function of $\ell$ for $(K_x',K_{\perp,\text{crit}})=(1.2,0.187)$ and $(1.4,0.49)$. We consider  values of  $\ell$ corresponding to partitions with  an even number of  rungs in an open chain.  Fitting our DMRG results using Eq. (\ref{entang}),  we obtain $c\approx 1.05$ and $1.03$, respectively. Considering different fitting intervals and system sizes, we have checked that our estimates are robust and the maximum deviation from $c=1$ is about $9\%$.  The logarithmic scaling of the entanglement entropy with the subsystem size   is   clear evidence of critical behavior at the transition. Moreover, our results show remarkable agreement with the central charge   predicted in Sec. \ref{sec:transition}. Finally,   we  have also investigated the effects of an explicit Klein-symmetry breaking by adding the interaction $K_\perp '\sum_r\tau^z_{r,1}\tau^z_{r,2}$ to the Hamiltonian  in Eq. (\ref{Heff}). While the values of the critical couplings shift with the perturbation,  no further transitions are observed and the central charge  remains the same. Therefore, the Klein symmetry does not affect the universality class of the transition.

\section{Future directions in 2D\label{sec2D}}
  
 The local pseudospin mapping of Eqs. (\ref{taux})-(\ref{rhoz2}) can   be used  to fabricate Hamiltonians with six-spin interactions that reduce to known spin-1/2 models in two dimensions once we freeze out  the $\boldsymbol \rho$ pseudospins. The phase with long-range order in $\tau^x$ would then correspond to a CSS. However, it is unclear if  this approach can  lead to deconfined transitions between chiral and magnetic phases in 2D  compass models. The existence of a robust continuous transition between competing ordered phases is conjectured to be connected to non-trivial symmetry properties of topological defects \cite{senthil2004deconfined,senthil2004quantum,levin2004deconfined,Wang2017}. Therefore, if the $\boldsymbol \tau$ pseudospin is defined 
envisioning the defects of $\tau^x$ and $\tau^z$ ordered phases on a given lattice, it may be possible to engineer a spin Hamiltonian in which gapping out the $\boldsymbol \rho$ pseudospin results in an effective model with a deconfined transition. An effective field theory description would then be described by a parton decomposition consistent with the defects \cite{Wang2017,jiang2019ising}.

A more interesting question is whether  a continuous phase transition from a CSS  to a collinear magnetic state or another   ordered phase  can be found  in models that do not require six-spin interactions.  Like the solvable model discussed in Sec. \ref{sec3}, the Yao-Kivelson model on the star  lattice \cite{Yao2007} exhibits spontaneous time-reversal-symmetry breaking and two chiral phases separated by a phase transition at which the gap for dynamical matter fermions closes. In this case, the  phases are topologically trivial and nontrivial chiral spin liquids distinguished by the Chern number.  In the exact solution using the Kitaev representation,  the topological excitations are vortices of  the emergent $\mathbb Z_2$ gauge field, which  bind Majorana zero modes in the nontrivial phase. One may then wonder if closing the vortex gap by  adding integrability-breaking perturbations to the Yao-Kivelson model  could drive an unconventional  transition to a magnetic phase. In a parallel development, the dynamics   of $\mathbb Z_2$ flux excitations and the relation  to phase transitions  in the extended Kitaev honeycomb model at zero magnetic field  has been   discussed based on  parton mean-field theories  \cite{Schaffer2012,Knolle2018}   and a variational approach   \cite{Zhang2021}.

Moving on to SU(2)-invariant  models, the situation becomes less clear. Here new dualities involving the scalar spin chirality \cite{Hikihara2003} may prove instrumental.  Numerically, a chiral spin liquid with  spontaneous time-reversal-symmetry breaking has been  found in the extended Heisenberg model on  the kagome lattice \cite{Gong2014,Gong2015}.    DMRG  results on cylinder geometries suggest that the quantum phase transition from the  chiral spin liquid  to  the  $q=(0,0)$ N\'eel state  is   at least not strongly first order     \cite{Gong2015}. The same can be said about transitions out of the chiral spin liquid phase in the triangular lattice Hubbard model \cite{Szasz2020}.


\section{Conclusions\label{secConclusion}}

In this work, we introduced a Kitaev-type model defined on the zigzag chain and showed the presence of two phases separated by a transition. On one side, we have a chiral spin state stabilized by coupling three-spin chiralities. On the other side, there is a collinear magnetic state also found in the natural extension of our model to two dimensions, the Kitaev model on the triangular lattice \cite{Maksimov2019}. Numerical analysis and field theory arguments indicate a continuous phase transition, which would be forbidden by the traditional LGW paradigm due to the competing nature of the order parameters. Furthermore, a low-energy parton construction suggests an emergent symmetry along with the condensation of topological defects (domain walls) in the transition, similar to the phenomenology found in deconfined quantum criticality  in two dimensions. Our work   then provides  an example of the recently found deconfined transitions in one dimension  \cite{mudry2019quantum,jiang2019ising,Roberts2021}. 

Further numerical investigation of this transition is also warranted. Critical exponents in correlation functions vary continuously  for a $c=1$ (Gaussian) transition, and it  would be interesting to see this behavior as one tunes  the microscopic parameters. Moreover,   our model  may host other phases and transitions for a different range of parameters. We   leave the complete mapping of the ground state phase diagram and the study of correlation functions at criticality for future work.

\begin{acknowledgments}
We thank J. C. Xavier for discussions on the DMRG implementation and the High-Performance Computing Center (NPAD) at UFRN for providing computational resources. We  acknowledge funding by  Brazilian agency CNPq (R.A.M. and R.G.P.). Research at IIP-UFRN is supported by Brazilian ministries MEC and MCTI. This work was also  supported by  a grant from the  Simons Foundation (Grant Number 884966, AF). 
\end{acknowledgments}

\bibliographystyle{apsrev4-2}       
\bibliography{ref}

\end{document}